\documentclass[letterpaper,twocolumn,10pt]{article}
\usepackage[T1]{fontenc}
\usepackage{amsthm}
\usepackage{array}
\newcolumntype{P}[1]{>{\centering\arraybackslash}p{#1}}
\usepackage{booktabs}
\usepackage{color}
\definecolor{lightgray}{rgb}{0.95, 0.95, 0.95}
\definecolor{darkgray}{rgb}{0.4, 0.4, 0.4}
\definecolor{purple}{rgb}{0.65, 0.12, 0.82}
\definecolor{ocherCode}{rgb}{1, 0.5, 0} 
\definecolor{blueCode}{rgb}{0, 0, 0.93} 
\definecolor{greenCode}{rgb}{0, 0.6, 0} 
\usepackage[bookmarks=false]{hyperref}
\hypersetup{
  colorlinks = true,
  linkcolor = blue,
  anchorcolor = blue,
  citecolor = blue,
  filecolor = blue,
  urlcolor = blue
}

\usepackage{times}

\usepackage{listings}
\usepackage{etoolbox}

\usepackage{geometry}
\usepackage{paralist}

\usepackage{subfig}

\usepackage{tikz}
\usepackage{pgfplots}

\theoremstyle{plain}
\newtheorem{thm}{Theorem} 

\theoremstyle{definition}
\newtheorem{definition}[thm]{Definition} 
\newtheorem{proposition}{Proposition} 

\usepackage{blindtext}
\usepackage[colorinlistoftodos,prependcaption,textsize=tiny]{todonotes}
\usepackage{regexpatch}
\makeatletter
\xpatchcmd{\@todo}{\setkeys{todonotes}{#1}}{\setkeys{todonotes}{inline,#1}}{}{}
\makeatother

\makeatletter
\lstdefinelanguage{HTML5}{
	sensitive=true,
	keywords={%
		typeof, new, true, false, catch, function, return, null, catch, switch, var, if, in, while, do, else, case, break,
		html, title, meta, style, head, body, script, canvas,
		border:, transform:, -moz-transform:, transition-duration:, transition-property:,
		transition-timing-function:
	},
	otherkeywords={<, >, \/},   
	ndkeywords={class, export, boolean, throw, implements, import, this},   
	comment=[l]{//},
	morecomment=[s]{/*}{*/},
	morecomment=[s]{<!}{>},
	morestring=[b]',
	morestring=[b]",    
	alsoletter={-},
	alsodigit={:}
}
\makeatother

\usepackage[%
  acronym, shortcuts,
  xindy={language=german-duden,codepage=utf8}
]{glossaries}

\renewcommand*{\CustomAcronymFields}{%
  name={\the\glsshorttok},
  description={\the\glslongtok},
}

\SetCustomStyle

\newcommand{\ie}{\textit{i.e.},}
\newcommand{\eg}{\textit{e.g.},}

\newcommand{\attacker}{\mathcal{A}}

\newacronym{AES}{AES}{Advanced Encryption Standard}
\newacronym{ABE}{ABE}{Attribute Based Encryption}
\newacronym{CSS}{CSS}{Cascading Style Sheets}
\newacronym[longplural={Content Security Policies}]{CSP}{CSP}{Content Security Policy}
\newacronym{DOM}{DOM}{Document Object Model}
\newacronym{XSS}{XSS}{Cross-Site Scripting}

\begin{document}
\date{}

\title {\Large \bf After You, Please: Browser Extensions Order Attacks and Countermeasures}

\author{
{\rm Pablo Picazo-Sanchez}\\
Chalmers $\mid$ University of Gothenburg, \\Gothenburg, Sweden
\and
{\rm Juan Tapiador}\\
Carlos III University \\Madrid, Spain
\and
{\rm Gerardo Schneider}\\
Chalmers $\mid$ University of Gothenburg, \\Gothenburg, Sweden
} 

\maketitle

\begin{abstract}
Browser extensions are small applications executed in the browser context that provide additional capabilities and enrich the user experience while surfing the web. The acceptance of extensions in current browsers is unquestionable. For instance, Chrome's official extension repository has more than 63,000 extensions, with some of them having more than 10M users. When installed, extensions are pushed into an internal queue within the browser. The order in which each extension executes depends on a number of factors, including their relative installation times. In this paper, we demonstrate how this order can be exploited by an unprivileged malicious extension (\ie\ one with no more permissions than those already assigned when accessing web content) to get access to any private information that other extensions have previously introduced. 
Our solution does not require modifying the core browser engine as it is implemented as another browser extension. We prove that our approach effectively protects the user against {\em usual} attackers (\ie\ any other installed extension) as well as against {\em strong} attackers having access to the effects of all installed extensions (\ie\ knowing who did what). We also prove soundness and robustness of our approach under reasonable assumptions.
\end{abstract}



\section{Introduction}\label{sec:introduction}

Web browsers have become essential tools that are installed on nearly all computers. The most popular browsers as of this writing (April 2018) are Chrome (77.9\%), Firefox (11.8\%), Internet Explorer/Edge (4.1\%), Safari (3.3\%) and Opera (1.5\%) \cite{w3schools}.
Most browsers allow users to install small applications, generally developed by third parties, that provide additional functionality or enhance the user experience while browsing. Such plug-ins are known as \textit{browser extensions} and they interact with the browser by sharing common resources such as tabs, cookies, HTML content or storage capabilities. 


In this paper we focus on Chromium \cite{chromium}, which is an open source browser and the basis for Chrome, Opera, Comodo, Dragon and the Yandex browser. Extensions installed in Chromium can also run in all mentioned browsers. The execution engine is exactly the same in all the browsers and follows the same pipeline model that will be explained in some detail later (Section \ref{sec:threatmodel}). For this reason, we will refer to Chrome and Chromium interchangeably. Extensions in Chromium can be of three types: \emph{content scripts}, \emph{background pages}, or both. In what follows, our main focus is on content scripts, which are JavaScript files that run in the context of the loaded web page. It is important to emphasize that the main aim of content scripts is to access and interact with the \gls*{DOM}. This fact alone raises a fundamental privacy question, since it is explicitly assumed that extensions will have full access to any (sensitive or not) content that the user is accessing. Browsers (including Chromium) dodge this issue by assuming that the user should trust the extension before installing it. In this paper we do not address this problem, which is essentially related to determining if an extension's behavior is benign or malicious.


When analyzing the security and privacy implications of browser extensions, one question that has been largely overlooked is the potential leakage of information among extensions. In nearly all browsers, each content script uses its own wrapper of the \gls*{DOM} to read and make changes to the page loaded by the browser. They also run in a dedicated sandbox that the browser provides for security reasons. However, there is no isolation in terms of privacy, since all changes an extension performs in its own \gls*{DOM} are automatically synchronized with the main \gls*{DOM}. One straightforward---but nonetheless important---consequence of this is that a malicious extension could eavesdrop on other extensions (\ie\ it can get access to the data they put on the \gls*{DOM} and observe their actions) and even manipulate their behavior by acting on their \gls*{DOM} elements (\eg\ clicking on elements introduced by another extension). An attacker can exploit this using two different strategies:
\begin{compactenum}
	\item \textbf{Exploiting the order.} The way Chromium manages extensions (see Section \ref{sec:threatmodel}) introduces a default execution order among extensions with undesirable consequences. One key issue is that the $n$-th extension in the pipeline can learn all contents introduced by the first $n-1$ extensions in the HTML document. Thus, extensions located at the end of the pipeline enjoy more privacy than ones installed earlier. 
	\item \textbf{The order-independent attacks.}
	Some of attacks enabled by the absence of effective isolation among extensions' actions do not require exploiting the execution order (\ie\ getting the malicious extension to be placed at the end of the execution pipeline). 
	However, exploiting the order provides the attacker with a privileged position that facilitates such attacks, which will result in a simpler code for the malicious extension that will increase the chances of passing the analysis performed by official stores \cite{Jagpal2015}. 
\end{compactenum}

This lack of effective isolation is not only inherent to Chromium's extension model, but also explicitly acknowledged. Browsers such as Chrome do not even attempt to guarantee some form of ``non-interference'' among extensions. On the contrary, developers are encouraged to implement appropriate mechanisms to protect any sensitive information that ends up in the \gls*{DOM}, since it is assumed that other extensions could simply read or manipulate it. Even if browsers do not factor this into their threat model, we believe that this is a serious vulnerability that has not been discussed before. More importantly, it can be easily exploited by a malicious extension, regardless of the fact that it is explicitly assumed in the browser's extension model or not.

\begin{figure}
	\centering
	\includegraphics[width=\linewidth]{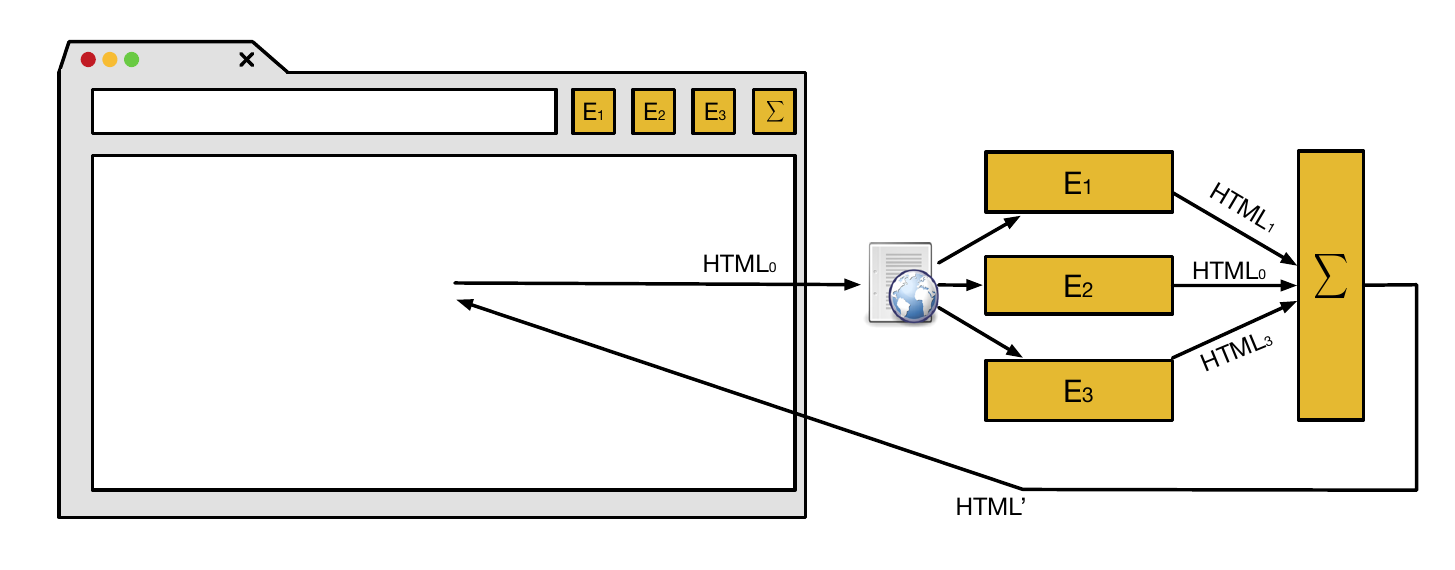}
	\caption{Modified extension execution pipeline in our solution.}
	\label{fig:proposed_solution}
\end{figure}

We discuss a vulnerability inherent to the way extensions are handled in Chromium, formalize this problem in terms of knowledge gained by the attacker and  propose a solution that provides practical security isolation among extensions and does not require altering the core browser engine. The key idea is to replace the extension pipeline by a (simulated) parallel execution model in which all extensions receive the same input page (see Fig.~\ref{fig:proposed_solution}). An additional component identifies the changes introduced by each extension and applies all of them to the original input page. We prove properties (soundness and robustness) of our solution and also discuss limitations of this approach.

\section{Chrome Browser Extensions}\label{sec:browserextensions}

\noindent{\textbf{Events order in JavaScript.}}\label{sec:JavaScriptEvents}
In JavaScript, the {\em event propagation} mechanism determines in which order an event is received by HTML elements. For instance, when two nested elements are subscribed to the same event 
and this event is fired, there are basically two ways to propagate the event in the \gls*{DOM}: {\em bubbling} and {\em capturing}. By using capturing, the event is initially handled by the root element and propagated to the its children. In contrast, with bubbling the event is initially captured and handled by the children (leaves nodes in the \gls*{DOM} tree) and then propagated to their parents.


In JavaScript, extensions subscribe to events by using the \path{addEvenListener()} function. In our example above we use capturing, so the first \textit{alert()} will correspond to the \path{<div id="1">}. In case of using bubbling, then the first \path{alert()} will correspond to the \path{<div id="2">} element.

Apart from the JavaScript event propagation mechanism, extensions developers can define one additional {\em order level} through a property named \path{run_at} in the \path{manifest.json} file that allows them to control at which moment the extension will be injected. That property might be: \path{document_start}, \path{document_end} or \path{document_idle}. 
When the value is \path{document_start}, the extension is injected when the document element is created. Setting it to \path{document_end} would cause the extension to be injected when the \gls*{DOM} is completed but before any other subresources are loaded, such as \eg\ images, iframes, etc. Internally, Chromium loads the extension when the \path{DOMContentLoaded()} is triggered. 
Finally, the \path{document_idle} value would cause the extension to be injected after \path{document_start} and before \path{document_end}, that is, once the page has been created and after the \gls*{DOM} is loaded. Internally, Chromium loads the extension after the trigger \path{window.onload()} is fired.

Additionally, Chrome currently works by delegating to the HTML parser the way the content scripts are inserted when they are tagged as either \path{document_start} or \path{document_end}. Thus, if the HTML parser schedules \texttt{document\_start} or \path{document_end} as tasks, then content scripts are inserted in separate tasks. However, content scripts tagged as \path{document_idle} are always injected in separate tasks.

Despite of the existence of the aforementioned strategies to control the execution order of extensions, explicitly writing them does not unequivocally determine the order in which they will be executed. Whenever two or more extensions have the same configuration parameters, the Chrome extension engine decides which one will be executed based on the extensions' installation date. This behavior follows a FIFO policy: the oldest installed extension will be the first to be executed whereas the newest will be the last one to be executed.

Apart from the event management mechanism explained above, it is worth mentioning how Chrome manages tasks and microtasks. A task---a click event, for instance---is run in its own thread and is composed of a set of JavaScript sentences, actions to handle the event, which belong to the \textit{event loop}. All tasks are queued and executed sequentially. Moreover, when a \path{setTimeout} is used in the event loop, the callback function that is executed asynchronously is queued as a new task. This is specially useful for monitoring delayed functions in browser extensions.

Nevertheless, some operations can be also executed in the middle of a task execution, \eg\ to make something asynchronous without being scheduled as a new task and queued in the tasks queue. Those operations are called microtasks (composed of promises and mutation observers) and are executed intermediately after the task execution. The main reason for this new types of queues is to enrich the user experience. However, microtasks are not executed when the event loop is not empty, \eg\ if two click events are fired using JavaScript code and the function that handles the click event uses promises, those microtasks will not be run until both clicks events are executed. We refer the reader to \cite{microtasks} for more information and practical examples about how tasks, microtasks and how their execution queues work. In this work however, we are not taking microtasks into consideration and they are left as future work as it is mandatory to modify the Chromium's source code to take them under control.

\noindent{\textbf{Extensions in Chromium's source code.}}\label{sec:sourcecode}
The security model of browser extensions in Chromium is based on isolated worlds in (JavaScript) V8. Its main purpose is to isolate the execution of different untrusted content scripts (with a wrapper of the original \gls*{DOM}) while keeping the main \gls*{DOM} structure synchronized.

Essentially, a \textit{world} is a \textquotedblleft concept to sandbox \gls*{DOM} wrappers among content scripts\textquotedblright~\cite{csworlds}. Each world has its own \gls*{DOM} wrapper, yet there might be different instances from one particular world and, thus, all of them would share the same Blink C++ \gls*{DOM} object. The main reason for this partition is that instances belonging to the same world cannot share \glspl*{DOM} but can share C++ \gls*{DOM} objects, \ie\ no JavaScripts can be shared between different worlds but C++ \gls*{DOM} objects can be, thus permitting to run untrusted content scripts on shared \gls*{DOM}. 

Roughly speaking, in terms of browser extensions this world concept means that the content scripts of each extension will run its own JavaScripts over different \glspl*{DOM}. However, all these \glspl*{DOM} are synchronized so that all changes made by each individual JavaScript will automatically be sent to other \glspl*{DOM} (other wrappers and the main \gls*{DOM} the user sees).

\begin{figure}
	\centering
	\includegraphics[width=\linewidth]{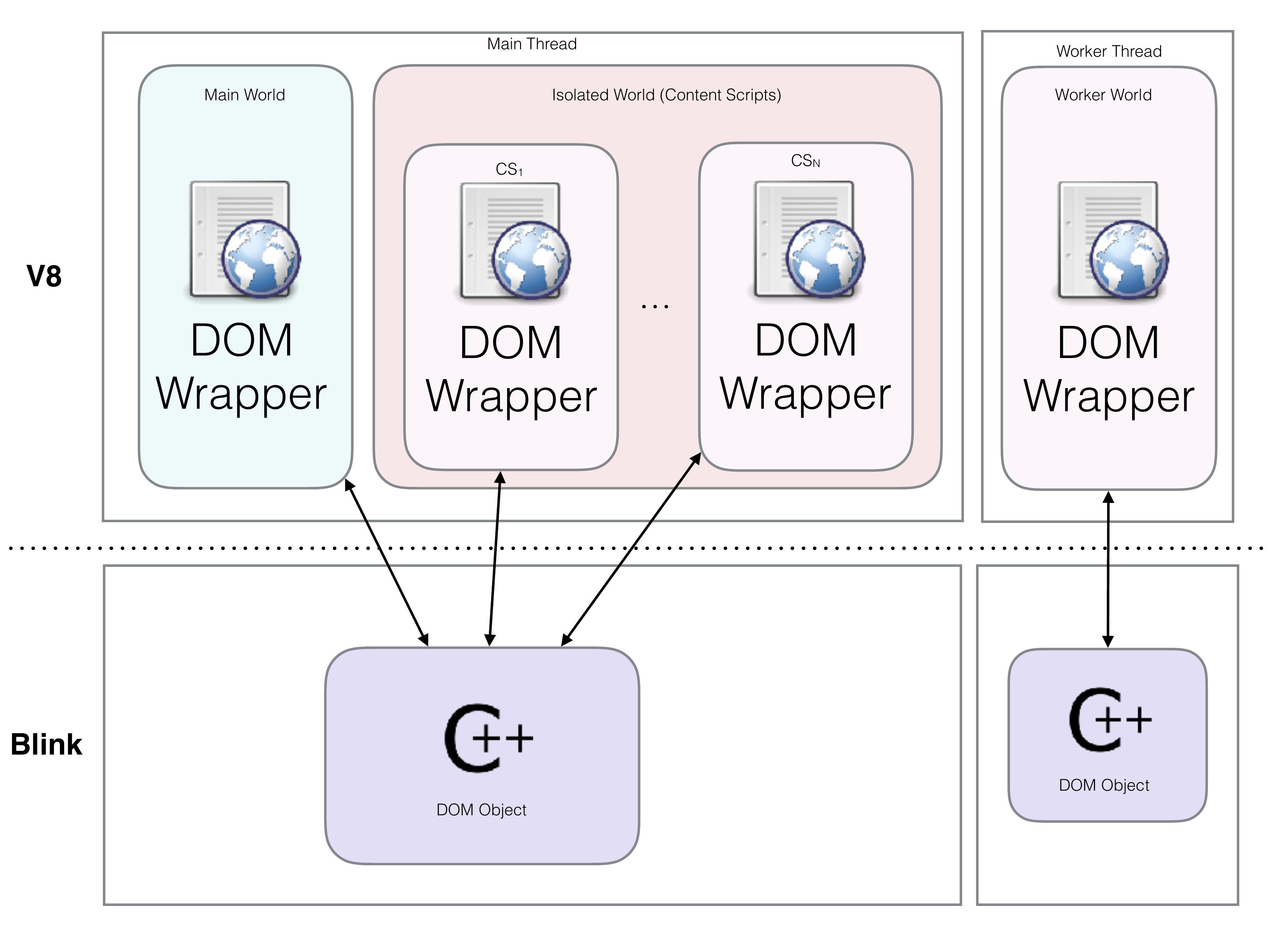}
	\caption{Browser Extensions Architecture in Chromium}
	\label{fig:extensionsarchitecture}
\end{figure}

According to the official documentation, V8 has three different \textit{worlds}: a main world, an isolated world and a worker world. A main world is where the original \gls*{DOM} with all its original scripts are executed. An isolated world is where the content scripts of the extensions are executed---all of them can access the main \gls*{DOM} through Blink C++ shared objects. Finally, a worker world is associated with threads in such a way that each isolated world is associated with one worker, \ie\ the main thread is the main world plus each of the content scripts. Figure \ref{fig:extensionsarchitecture} represents how Chrome manages and isolates content scripts of browser extensions.

Overall, this isolation mechanism prevents Chrome from being vulnerable to attacks such as the one recently demonstrated in \cite{Saini2016} against Firefox, whose security model lacks isolation. Nevertheless, Chrome's security model has not considered privacy between extensions as part of its architecture. This allows, for instance, that if the Pinterest extension inserts a \path{<span>} element on each picture contained in the \gls*{DOM}, then all these changes will automatically be visible to the rest of the extensions, regardless of whether they run in isolated worlds. This observation is the basis for the attack discussed in this paper.

\section{Attacker Model}\label{sec:threatmodel}



%
%

Chromium manages all installed extensions through the class \textit{ExtensionRegistry} which is implemented in \textit{extension\_registry.cc} and the associated header file \textit{extension\_registry.h}. This class implements methods to add, remove or retrieve all extensions that for a particular browser context have been enabled, disabled, blocked, blacklisted, etc. Each set of extensions is internally managed through the \textit{ExtensionSet} class implemented in \textit{extension\_set.h} and \textit{extension\_set.cc}. This is just a standard C++ map to manage sets with methods to insert, remove and retrieve items. Importantly, it also provides a standard C++ iterator to enumerate all set elements. Overall, this means that installed extensions in Chromium are placed in a pipeline and are called sequentially to inject the content script(s) in the \gls*{DOM}. Apart from that, the position at which the browser injects content scripts is determined by a number of factors:
\begin{compactenum}
	\item First, by the implicit order declared in the \textit{run\_at} property in the \textit{manifest.json} file.
	\item If two or more extensions have the same \textit{run\_at} value, the browser tries to determine their execution order using the JavaScript event propagation mechanism \cite{capturing_javascript}.
	\item Finally, if the event propagation order is the same, extensions are executed according to their installation time: $A$ executes before $B$ if $B$ was installed after $A$.
\end{compactenum}
This pipeline works as 
follows. The content script of the first extension is inserted and then it is executed in an isolated world by using the method \textit{executeScriptInIsolatedWorld} (see \cite{csworlds} for more details about worlds in Chrome). The output of an extension is automatically synchronized with the shared \gls*{DOM}, so when the next extension is executed its wrapper \gls*{DOM} already contains all the changes that previous extensions have made.


\subsection{Manipulating the execution pipeline}
We assume that the attacker gets one malicious extension at the end of the execution pipeline. 
Even if the extension is not the last to be installed, Chromium's extensions model provides various mechanisms that an attacker can exploit to modify the order in which parts of the extension code will run. For example, a malicious extension can be marked to run once the \gls*{DOM} is loaded. This is achieved by setting the \textit{run\_at} property to \textit{document\_end} in the \textit{manifest.json} file. Additionally, the extension can use the \textit{capturing} JavaScript event propagation property to force that the fired event would execute the extension in the return journey of the event propagation (check \cite{capturing_javascript} for further details about this). Moreover, modifying the default execution order could be done by following two different approaches: 

\begin{compactenum}
	\item Through another extension's \textit{management} permissions, similarly to what the \textit{Extensity} extension does \cite{extensity}. Essentially, this extension enables and disables extensions automatically in the browser. The attacker will disable all installed extensions and then re-enable them again, but putting the malicious extension at the end. For this to work in practice, the user must explicitly approve the malicious extension requirement to extend its permissions (namely, management). 
	Since many users do not pay attention to requested permissions, this could guarantee success for the attacker.
	
	\item Modifying the \textit{Secure Preferences} file. This is a JSON configuration file that was initially thought to be modified only by the browser \cite{chrome_external}. However, Chrome allows developers to distribute extensions as part of other software so this file could also be externally modified by other processes. This is the basis for most of the malware installed in the browser because of its deficient security \cite{secure_preferences_chromium}. 
	The attacker can thus modify the \textit{install\_time} property in the \textit{Secure Preferences} file, and put her extension at the end of the pipeline. 
	See \cite{secure_preferences} for a detailed explanation on how to modify the manifest file.
\end{compactenum}

\subsection{Attack examples}\label{sec:examples}

To exemplify the problem, we tested the attack in a Chrome browser with four extensions already installed: Pinterest, Evernote, vidIQ Vision for YouTube and our custom extension. Our malicious extension subscribes to all possible JavaScript events in the browser (\ie\ in the web page context). This guarantees that the extension will always be executed whenever any event is fired.

The official extension of Pinterest, which has more than 10M users, parses the entire content and adds hidden \textit{<span>} elements on each picture it finds in the \gls*{DOM}, as well as some \gls*{CSS} elements. When the user triggers the \textit{onmouseover()} event by passing the mouse pointer over a picture, the span becomes visible in the form of a button and, if the user clicks on it, the picture is automatically shared on her Pinterest board. Assume now that the user has some ``secret boards'' defined in her Pinterest account to avoid sharing pictures with all the world. Our malicious extension can carry out the following actions:
\begin{compactenum}
	\item It can add a listener to the same \textit{onclick()} event to know which pictures the user adds to her account, and thus the photos will no longer be private.
	\item It can learn what pictures the user likes and it could share that information to an advertisement company \cite{Xing2015}.
	\item It could generate a \textit{click} JavaScript event on each picture, automatically sharing all pictures in the user's account without any confirmation pop-up.
	\item It can replace the picture the user wants to share by another one.
\end{compactenum}

Evernote Web Clipper has currently more than 4.5M users. The extension parses the web page and inserts some \gls*{CSS} code and hidden \textit{<span>} elements on each picture contained in the \gls*{DOM}. Additionally, it adds a contextual menu when the user performs a right click either on a single tag or in the whole document. Using this contextual menu, the user can add items such as meetings, personal notes, or any other information to her calendar. Our malicious extension can subscribe to the click events and, in addition to the attacks described for the Pinterest scenario, it could also learn all details about the notes or calendar entries added by the user. 

Finally, we tested it againts vidIQ Vision for YouTube. This extension has more than 500,000 users. Among other actions, it inserts a \textit{<div>} element in the right banner of the screen when a user visits Youtube in order to provide her with richer information and track her viewed videos. When the user visits Youtube for the first time, this extension asks her for her username and password (as a matter of fact, all extensions subscribed to either \textit{onkeydown()}, \textit{onkeypress()} or \textit{onkeyup()} events may get both the username and password). Our malicious extension, apart from getting the username and password, could also get all viewed videos and profile the user's habits.


\subsection{Modeling extension effects}

Before formally defining our attacker model, we first introduce some notation and definitions. In what follows, $E=\langle E_1, \dots, E_i, \ldots,$ $E_n\rangle$ ($n > 0$)\footnote{All the discussion below assumes that there is at least one extension installed.} 
will denote the ``set'' of extensions already installed in the browser, where the index indicates their default execution order (\ie\ $E_1$ is the first to be executed).

When extensions are executed, they have an \textit{effect}. For our purposes, we split such effects into two parts: a {\em functional} effect that is reflected on the changes done to the \gls*{DOM} the extension acts on, and some {\em side-effects} that are not directly reflected in the \gls*{DOM} (\eg\ sending information to other servers, interacting with the browser, executing external scripts, etc.). The functional effect of an extension $E_i$ when applied to a \gls*{DOM} will be denoted by $f_i$(DOM) = DOM$_i$. In this paper we are only concerned about the functional effect of \glspl*{DOM}, so all the results that follow only apply to what extension can do on the \glspl*{DOM} and, thus, no claim is done concerning extensions' side-effects.

Extensions can perform four different types of high-level operations while being executed: {\em insertions}, {\em deletions}, {\em updates}, and simply doing {\em nothing}. 

\begin{definition}\label{execution_pipeline}
	Let $E=\langle E_1, \dots, E_i, \ldots, E_n\rangle$ with $n>0$, be the set of extensions that a browser has already installed and \gls*{DOM}$_0$ the original content provided as input. We define the \emph{execution pipeline} as the result of the execution of  the $n$ extensions as composite functions: $f_n\circ 	\ldots \circ f_1$(DOM$_0$) = DOM$_n$.
\end{definition}	

\glspl*{DOM} can be seen as trees \cite{W3cDOM}. We will use this fact to define the above operations in terms of operations on trees. Thus, if extension $E_i$ only inserts elements in the \gls*{DOM}, then $f_i$(DOM) = DOM$_i$ where \gls*{DOM} is a subtree of \gls*{DOM}$_i$. In case $E_i$ only deletes something from \gls*{DOM}, then $f_i(DOM) = DOM_i$, where DOM$_i$ is a subtree of \gls*{DOM}. Finally, if $E_i$ only updates \gls*{DOM}, then $f_i$(DOM) = DOM$_i$ where \gls*{DOM} is equal to \gls*{DOM}$_i$ except for the field that has been updated.\footnote{Note that, to avoid over-formalization, we are not giving formal definitions for these operations in terms of trees as they are rather intuitive.} 

We assume a tree operation that allows us to compare \glspl*{DOM} and give us the {\em difference} between them: DOM - DOM'. 
Moreover, we say that \gls*{DOM} is {\em smaller or equal} than \gls*{DOM}' (denoted \gls*{DOM} $\leq$ \gls*{DOM}') if and only if \gls*{DOM} is a subtree of \gls*{DOM}'.\footnote{We define $<$, $>$, and $\geq$ as expected.}

Finally, we say that the {\em default knowledge} of an extension is the amount of information it can get from the \gls*{DOM} at the moment of its execution. Note that the actual knowledge of an extension might not be equal to the default knowledge. 


\begin{definition}\label{knowledge}\label{def:knowledge}
	Let $E=\langle E_1, \dots, E_i, \ldots, E_n\rangle$ with $n>0$ be as before,  \gls*{DOM}$_0$ the original content, and $EV=\{ev_1, ev_2, \ldots, ev_n \}$ the set of events that DOM$_0$ can fire. We define the \textit{default knowledge of an extension} 
	$E_i$ as $K(E_i)=\{ $DOM$_0 \mid \exists e \in EV \cdot e = load() \} \cup \{ $DOM$_{i-1} \}$ for $1< i \leq n$, and $K(E_i)=\{ $DOM$_0\}$ if $i=1$.     
	We say that an extension  $E_j$ {\em knows at least as much as} extension $E_i$ ($1\leq i,j \leq n$) if and only if $K(E_i) \subseteq K(E_j)$.
	We also define an order between extensions concerning the DOM they have direct access to: $E_{i} \sqsubseteq E_j$, if and only if the resulting \gls*{DOM} after executing both extensions is such that DOM$_{i} \leq $DOM$_j$.
\end{definition}	
Note that the {\em real} knowledge of an extension might not be equal (and neither a subset nor a superset) of the default knowledge. The reason is that, as we will see, this knowledge might be affected by attacks or by a solution to those attacks. The concept is in any case useful as it characterises what the extension knows by default, if no external interference is added to the expected behavior of how the browser works. 
On the one hand, if we have an execution pipeline such that the functional effect of all of them is only insertions or doing nothing. We say that the execution pipeline is \textit{monotonic with respect to the structure of the \gls*{DOM}} (or simply, that it is {\em monotonic}). We have then the following proposition concerning monotonic execution (sub)pipelines. The proof is trivial.
\begin{proposition}\label{prop:monotone}
	Given a sequence of extensions $\langle E_1$, $\dots$, $E_j, \dots,$ 
	$ E_h,\ldots,  E_n\rangle$  (with $n>0$) such that the subsequence $\langle E_j, \dots, E_h\rangle$ is monotonic, then $E_{i-1} \sqsubseteq E_i$ for $j < i \leq h$.
\end{proposition}

If an execution pipeline is such that the overall functional effect of all extensions is only insertions or doing nothing, we say that the execution pipeline is \textit{monotonic with respect to the structure of the \gls*{DOM}} (or simply, that it is {\em monotonic}). Conversely, if any extension $E_i$ in the execution pipeline deletes or updates information, then it is generally impossible to make any statement about whether any other extension knows more or less than $E_i$. For instance, an extension $E_{i-1}$ could delete something while extension $E_i$ adds it back, in which case any other extension $E_j$ ($j>i$) will not be able to detect that there has been a deletion in the past.

\subsection{Attacker model}
We consider two different types of attackers: {\em strong} and {\em usual} attackers. 
Intuitively, a \textit{strong attacker} is a malicious extension that has access to the output of all executions in the pipeline. Note that this provides the attacker not only with the effect of all extensions, but also with knowledge about which extension did what. Alternatively, a \textit{usual attacker} is a browser extension that only has access to the corresponding \gls*{DOM} that the extension receives as input when it is executed (plus the original \gls*{DOM}). More formally:

\begin{definition}
	A \emph{strong attacker} ($\attacker_s$) is an extension $E_{\attacker_s}$ that is interleaved in the execution pipeline such that $f_{\attacker_s} \circ f_n \circ f_{\attacker_s} \circ \cdots \circ f_{\attacker_s} \circ f_1 \circ f_{\attacker_s}(DOM) = DOM_n$. 
	This is the strongest attacker because it can know all the changes that all extensions have performed.
	A \emph{usual attacker} ($\attacker_u$) is an extension $E_{\attacker_u}$ that is executed in the $j-$th position of the  pipeline ($ j \leq n$) such that $f_n \circ \ldots$ $\circ f_{\attacker_u} \circ \cdots$  $\circ f_1(DOM_0) = DOM_n$, having the default knowledge any other extension in position $j$ could have. Note that $j>1$ as otherwise the attacker would learn nothing.
\end{definition}

A strong attacker has definitively more knowledge than any other in the pipeline and can thus take advantage of that. Note that, in particular, a strong attacker gets to know which extension did what changes since it can calculate the effect of each extension. The usual attacker can only infer partial information about the other extensions by diffing \gls*{DOM}$_0$ and the \gls*{DOM}$_{\attacker_u}$ that it receives as input. However, this attacker will know neither the number of extensions nor which operations they have performed over the content. Note that the gain of knowledge is not much over previous extensions except if \gls*{DOM}$_{\attacker_u}$ is part of a monotonic subsequence (cf.~Proposition \ref{prop:monotone}).

An interesting consequence of our threat model is that all extensions which are installed on the browser are potential usual attackers because they might have access to the original \gls*{DOM} and to the input \gls*{DOM} received from the previous extension in the execution pipeline (cf.~Definition \ref{def:knowledge}).

Remember that despite our proposed attack might be performed without exploiting the order, \ie\ a malicious extension could subscribe to all possible events in the \gls*{DOM}, the amount of needed source code to tackle all possible privacy attacks would be incredibly huge and infeasible due to the amount of possible extensions and attacks. 

In this work, we remark the existence of this security threat which is transparent even for the automatic static analysis of the source code that official repositories perform \cite{Jagpal2015}. Notice that by using our attack, the simplest dummy extension installed just after, for instance the official Pinterest extension, would detect the existence of the former one and thus, it can communicate to an external server to retrieve the customized exploit performing thus an adaptive attack.

Additionally, our scenario can handle situations where, two browser extensions developed by the same person or company but placed for instance at the beginning and at the end of the execution queue will actually access to different information and thus, collaborate to perform attacks like browser hijacking \cite{Xing2015,Rogowski2017}, or fingerprinting \cite{sjosten2018,Laperdrix2019} attacks.

\section{Our solution}\label{sec:approach}



\begin{figure}
	\centering
	\includegraphics[width=1\linewidth]{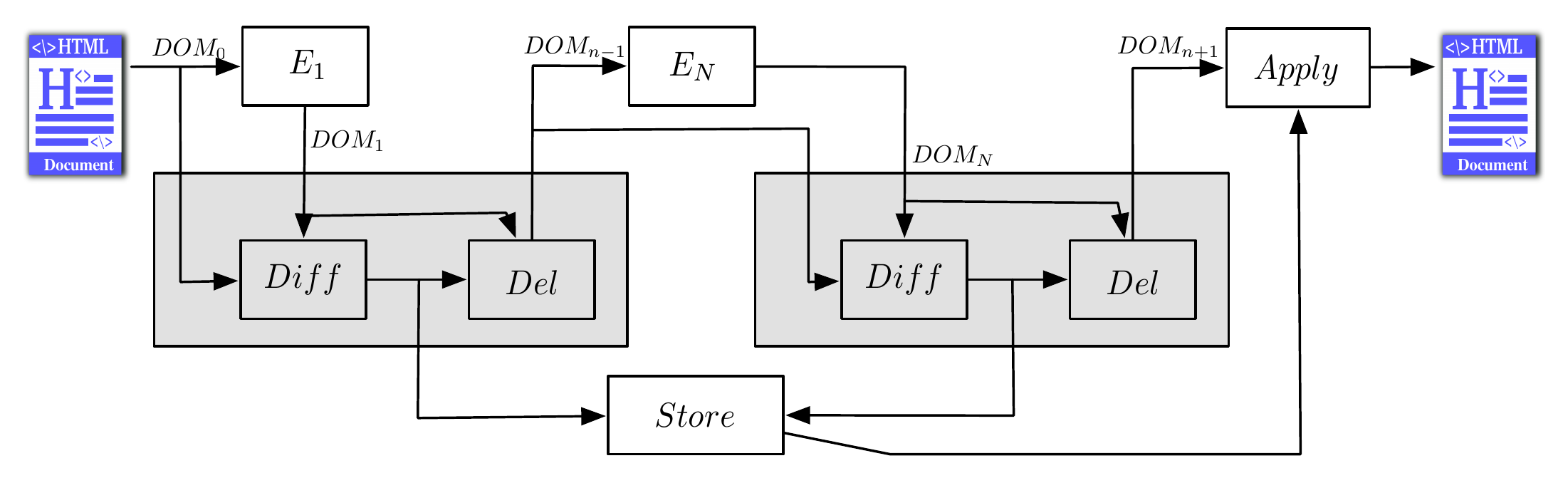}
	\caption{Architecture of our solution and its four main modules.}
	\label{fig:general_solution}
\end{figure}

Our solution introduces the notion of a \textit{monitor extension}, whose goal is to prevent regular extensions from learning from each other.
Intuitively, monitor extensions are used to detect all changes that an extension makes; log those modifications; delete them from the \gls*{DOM} passed on to the next extension in the execution pipeline; and, at the end of the pipeline, merge all changes to produce a final \gls*{DOM}. Figure \ref{fig:general_solution} shows the four main components of our scheme:
\begin{compactitem}
	\item The {\em Diff} module takes a pair of \glspl*{DOM} (namely, (\gls*{DOM}$_{i-1}$, \gls*{DOM}$_{i}$) and performs the difference between them (Diff=DOM$_{i}$ - DOM$_{i-1}$).
	
	\item The {\em Store} module is shared between all monitor extensions and collects all changes in a table. This table can be seen as a \textit{patches} table with the following format: <\textit{Operation}>, <\textit{Position}>, <\textit{Action}>. 
	
	\item The \textit{Del} module removes all changes from DOM$_{i}$, that is, DOM$_{i}$ = DOM$_{i-1}$ - Diff.
	
	\item The \textit{Apply} module, which is placed at the end of the pipeline, takes all stored differences and patches the \gls*{DOM} by applying them in order.
\end{compactitem}

Note that our solution could be simplified by removing the \textit{Del} method. Thus, once the difference has been computed, the \gls*{DOM} passed on to the next extension would be just the original \gls*{DOM} (\gls*{DOM}$_0$). However, by implementing a \textit{Del} module, our approach is more general since it allows to introduce some policies to share limited amounts of information among extensions. This point, however, is not further explored in this work.

More formally, this solution implements the following transformation over the input \gls*{DOM} :  $f_{E_{final}} \circ f_{E_n}$ $\circ f_{E_{moni_{n-1}}} $ $\circ \ldots \circ f_{E_{moni_1}} $ $\circ f_{E_1} $ $\circ f_{E_{initial}}($DOM$_0$) = DOM$_n$. The information flow is as follows. Assume that Alice accesses a web page. The browser requests the URL and, once the \gls*{DOM}$_0$ tree is retrieved, the first isolated world corresponding to the initial extension ($E_{initial}$) is executed. This first extension is not part of the general solution (see Figure \ref{fig:general_solution}), but we found out that, when we tried to implement it in a real setting, it is needed because Chrome---and other browsers in general---do some pre-processing to the \gls*{DOM} (\eg\ closing forgotten open HTML tags, adding some mandatory HTML tags or changing everything into lower case). This initial extension does not add any changes to the \gls*{DOM}. We note that an extension can request the same content directly by using the \textit{XMLHttpRequest} JavaScript object, but the received \gls*{DOM} could be completely different from the current \gls*{DOM} because of that browser pre-processing. 

After that, the output of the initial extension \gls*{DOM}$_0$ and the rest of the \glspl*{DOM} wrappers are synchronized. At this point, the first official extension is run and may perform some actions over the content. The resulting (\gls*{DOM}$_1$) is the input to the next \textit{monitor extension}, plus the initial \gls*{DOM} (\gls*{DOM}$_0$) needed to get the difference between both \glspl*{DOM}: Diff=DOM$_1$-DOM$_0$. All the possible resulting values of this operation are stored ($Store$) for the final post-processing ({\it patch} operation), and the difference Diff is then removed from the output of the extension \gls*{DOM}$_1$. It is worth noting that this new \gls*{DOM} will be equal to the original \gls*{DOM} in most cases, \ie\ our solution will be valid whenever the execution pipeline follows a monotonic sequence. This process is repeated until the last official extension eventually produces the final \gls*{DOM}$_{n}$ output. This last \gls*{DOM} is then provided as input to the final extension ($E_{final}$), which will take all stored changes and will apply them to the \gls*{DOM}$_{n}$, thus generating the final document. 


\subsection{Advantages, properties and limitations of our approach}

%

\noindent\textbf{How intrusive is our solution?} That is, how much of the extensions' (good and expected) behavior do we modify while achieving our goal of preserving privacy? Our solution always preserves the behavior of the original browser execution model (\ie\ the final output with or without our solution is exactly the same).  In some sense, we do want to make sure that the order of execution is irrelevant with respect to the knowledge the extensions {\em should} get (\ie\ not accessing sensitive information they are not allowed to as an effect of this information being {\em passed} by other extensions), but we also know that the outcome of the executions of such extensions might be modified by our approach eventually modifying some of the expected output. 
Let us consider an example showing the possible effects of our solution. Let $E_i$ be an extension that changes the \gls*{DOM}'s background color to black (let us assume the original color was white and that there is text both in black and blue). Let us consider that a later extension in the execution pipeline, $E_j$ ($1 \leq, i < j \leq n$) changes the background color to white. It is clear that in the current order, the final outcome is that the \gls*{DOM}'s background color is white and all the text is readable. 
That being said, it is clear that in case the extensions were executed in different order (first $E_j$ and then $E_i$) the outcome will be very different: not only the background will be black (instead of white), which by itself does not seem to be a big deal, but more importantly there will be some text not visible to the user. This not only affects the usability of the \gls*{DOM} (the black background will hide all the black text so the user will not be able to see it), but may introduce some security issues (the hidden text might be clicked accidentally producing undesired effects).

This is however, an inherent behavior of the browser and our proposed solution does not modify the default behavior of the browser, \ie\ a given HTML content looks the same with a set of extensions enabled and with the same set and the proposed solution. Moreover, JavaScript periodical tasks such as \textit{setInterval(callback, delay)} are not covered in detail with our solution. This method automatically enqueues the function defined in the callback in the task queue. For instance, if the extension A uses this method to get all password fields from the page the user is visiting each 2 seconds. This is a completely different scenario because the execution of this task cannot be controlled through JavaScript code alone. 

\noindent\textbf{Does our solution indeed mitigate possible attacks?} According to the definitions given in Section \ref{sec:threatmodel}, the knowledge of an extension executed in position $j$ ($1 < j \leq n$) is the same knowledge as the previous extension ($E_{j-1}$) in the pipeline plus the actions that $E_{j-1}$ performs over the \gls*{DOM} ($DOM_{j-1} \cup DOM_j $). On the contrary, when we measure the knowledge of an extension with our solution, it is thus decreased to $DOM_0$ (given that our solution only passes the original DOM to each extension).
Our solution also mitigates a {\em strong} attacker by limiting what she gets to know in the same way as for the {\em usual} attacker: our interleaving guarantees that a {\em strong} attacker only gets to know the original \gls*{DOM}. The reduction in knowledge is of course more significant than in the {\em usual} attacker (Section \ref{sec:threatmodel}).

\noindent\textbf{How robust is the approach?} That is, can we guarantee that a {\em strong} attacker cannot bypass our solution? One may think that a {\em strong} attacker could attack our solution by interleaving extensions between our monitor extensions (before and after) thus bypassing our protection in order to get access to the effects of the installed extensions before being modified by our monitor extensions, and then restoring it after our modification. To do so, the attacker must create an extension with the \textit{management} privileges. That, however, would only be possible if the user explicitly grants that permission to the attacker. The best we can do is then to show a warning message to the user as soon as we detect the presence of such malicious extension and rely on that the user blocks the attacker. If the user grants the permission, we are thus vulnerable to the attack.
In order for our proposed solution to be able to detect the presence of such attacks 
our extension would need to have \textit{management} privileges. This could only be granted by the user at installation time. 

\begin{proposition}
	Our extension-based solution is robust against {\em strong} attackers under the assumption that our (initial, middle and final) monitor extensions are given \textit{management} privileges, and that the user does not explicitly give \textit{management} privileges to the attacker. 
\end{proposition}

In case the user (accidentally or consciously) gives the needed privileges for a {\em strong} attacker to install his extensions, our solution would be able to detect that and communicate it to the user. Indeed, a {\em strong} attacker would need to install $n+1$ extensions interleaved between any two extensions, and our monitor extensions would be able to detect that. So, we have a way to detect this issue, notify it, and ask the user to uninstall the extension. Besides, by identifying this we would be able to keep a black list of malicious extensions.

Finally, it is worth mentioning that most of the aforementioned questions would be solved by modifying the Chromium's source code. Note that an attacker might insert as many extensions as desired and could even alter the execution order. By modifying the source code, all extensions receive a fresh copy of the original HTML and, thus, no-one will learn about the actions executed by other extensions. This solution uses a similar approach but requires modifying (and recompiling) the core of Chromium to achieve a real isolated execution. At a logical level, it works exactly the same as the general solution depicted in Figure \ref{fig:general_solution}. 

\section{Experimental results}\label{sec:results}

We have studied the following performance indicators according to \cite{MeasuringRendering} as well as the W3C consortium \cite{W3CTiming}: 
\begin{inparaenum}
	\item memory consumption;
	\item time needed to parse the HTML;
	\item when the \textit{onLoad} event is fired (many JavaScript files wait for this event);
	\item the \textit{processing time} which means that all resources have been loaded (\gls*{DOM} is completed \ie\ the loading spinner has stopped spinning), and;
	\item a final test to show the total time that Chrome needs to generate the \textit{onLoad} event, \ie\ the page is ready.
\end{inparaenum}
All the experiments but the memory consumption were carried out accessing the Alexa's Top 30 web sites and averaging the results over 50 runs. Additionally, in order to measure all the time-based metrics, we have used the DevTools profiling tools provided by the browser.

Our extension based solution inserts a middle monitor extension between every two original extensions, plus the initial and the final ones. Thus, the number of total extensions is $2n+1$ ($n$ original extensions plus $n+1$ added monitor extensions, including the initial and final ones). In order to test what impact these additions have on both Chrome's performance and the user experience, we have installed a set of original extensions in a MacBook Air with 2.2 GHz Intel Core i7 CPU and 8 Gb of RAM. The Chrome version where all test have been run is 60.0.3112.78 (Build official) (64 bits). We used the 10 most downloaded browser extensions from the Chrome Web Store, since according to \cite{addonsstats}, the average number of installed extensions per user is 5. 

All figures related to the monitor extensions depend on the number of original extensions installed in the browser ($2n+1$). In our experiments, the number of extensions is related to the original extensions installed. This number varies if the experiment is performed by using the original extensions or our proposed solution. For instance, when we say that with 5 extensions it takes 1.3 seconds to load all the scripts of a entire page, it means that in reality there are 11 extensions installed in the browser: 5 original extensions, plus 4 middle extensions, plus 1 final extension, plus 1 initial extension. On the contrary, 5 extensions on the \textit{original extension} experiment means that only the 5 original extensions are installed in the browser. Additionally, for all the experiments we have measured times without the browser's cache and by launching one new, fresh instance per experiment, \ie\ we have closed and opened Google Chrome each time we added a new browser extension to measure RAM consumption and user experience times.

\subsection{RAM Consumption}
To measure memory consumption, we have used the developer tools provided by Chrome. Table \ref{fig:stats} shows the impact on the browser performance in terms of RAM consumed in KB. We have isolated the execution of the original extensions and the monitor extensions in order to show that the impact of our proposed solution is almost negligible in comparison to the performance of the original extensions. Moreover, both the initial and the final extension consume 13 KB of RAM each, whereas our monitor extensions consume 11 KB of RAM on average. These extensions differ considerably from extensions such as AVG Web TuneUp, AdBlock or Ad Block Plus, which consume 27.6 KB, 190.3 KB, and 11.3 KB of RAM on average, respectively. Note that the size of such extensions depends in fact on the content of the web page. For instance, a page containing a substantial amount of advertisements would make Ad Block to consume much more memory. From the results we can conclude that the impact of our solution is approximately linear in the number of extensions. More concretely, our proposed solution decreases performance by a factor of 1.15 per installed extension in terms of RAM.

\begin{table}
	\centering
		\resizebox{\columnwidth}{!}{%
	\begin{tabular}{ | P{2cm}  P{1.8cm}  P{1.8cm} | P{2cm}  P{1.8cm}  P{1.8cm} |}
		\hline
		\#Extensions & Originally  & Solution & \#Extensions & Originally & Solution\\ 
		\hline 
		2 & 217,9Kb & 255,0Kb & 7 & 420,9Kb & 513,1Kb	\\ 
		3 & 331,6Kb & 379,7Kb & 8 & 492,0Kb & 595,2Kb	\\ 
		4 & 348,7Kb & 407,9Kb & 9 & 504,3Kb & 618,5Kb	\\ 
		5 & 374,1Kb & 444,2Kb & 10 & 527,1Kb & 652,3Kb	\\ 
		6 & 392,1Kb & 473,3Kb &     &     & 	\\ \hline
	\end{tabular}
	}
	\caption{RAM Consumption}\label{fig:stats}
\end{table}

\subsection{Impact on user experience}

\begin{figure*}
	\subfloat[{\small HTML Parse }\label{fig:htmlparse_time}]{%
		\resizebox{7.5cm}{!}{
			\includegraphics[]{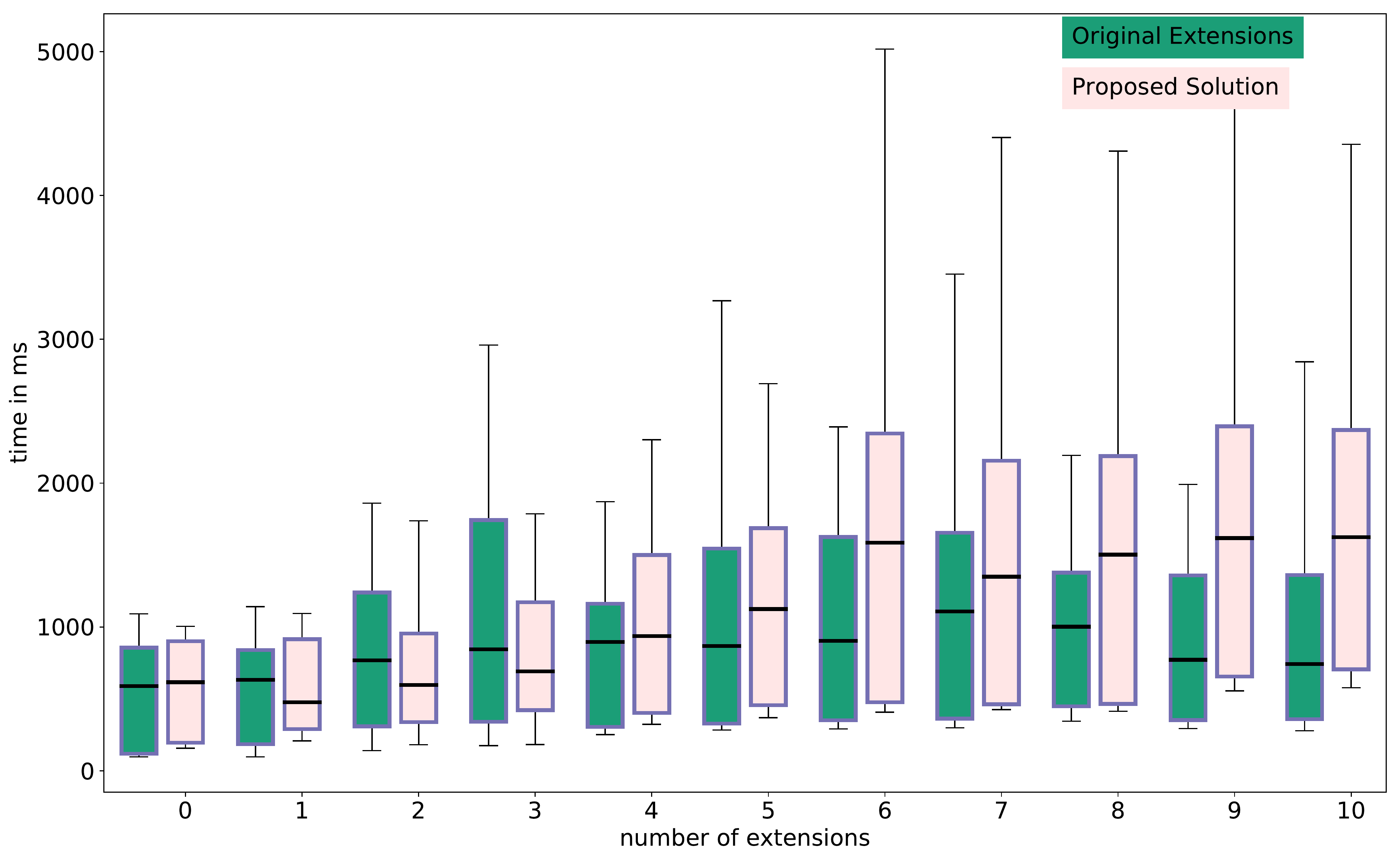}
		}
	}
	\subfloat[{\small onLoad}\label{fig:onload_time}]{%
		\resizebox{7.5cm}{!}{
			\includegraphics[]{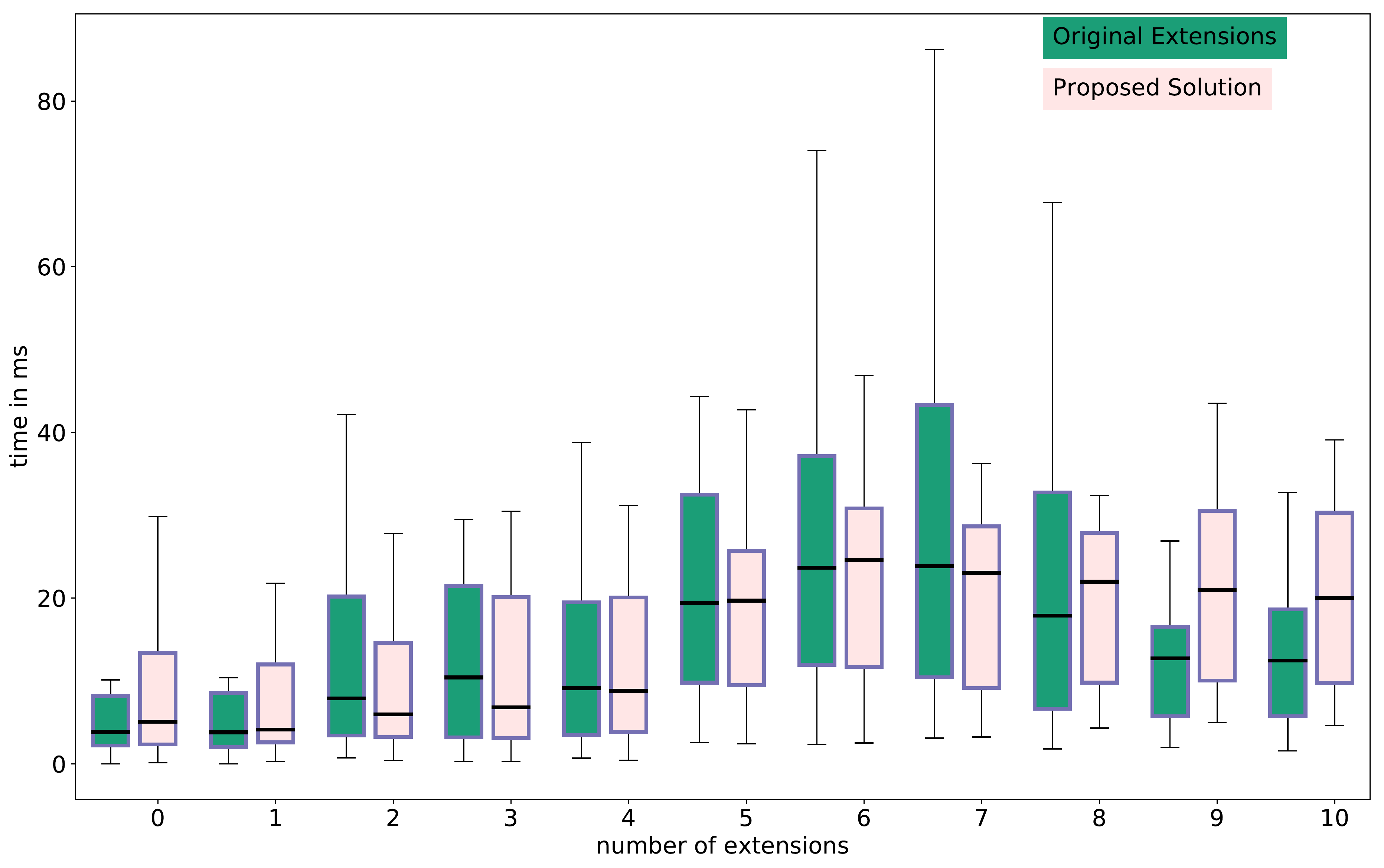}
		}
	}
	
	\subfloat[{\small Processing Time}\label{fig:processing_time}]{%
		\resizebox{7.5cm}{!}{
			\includegraphics[]{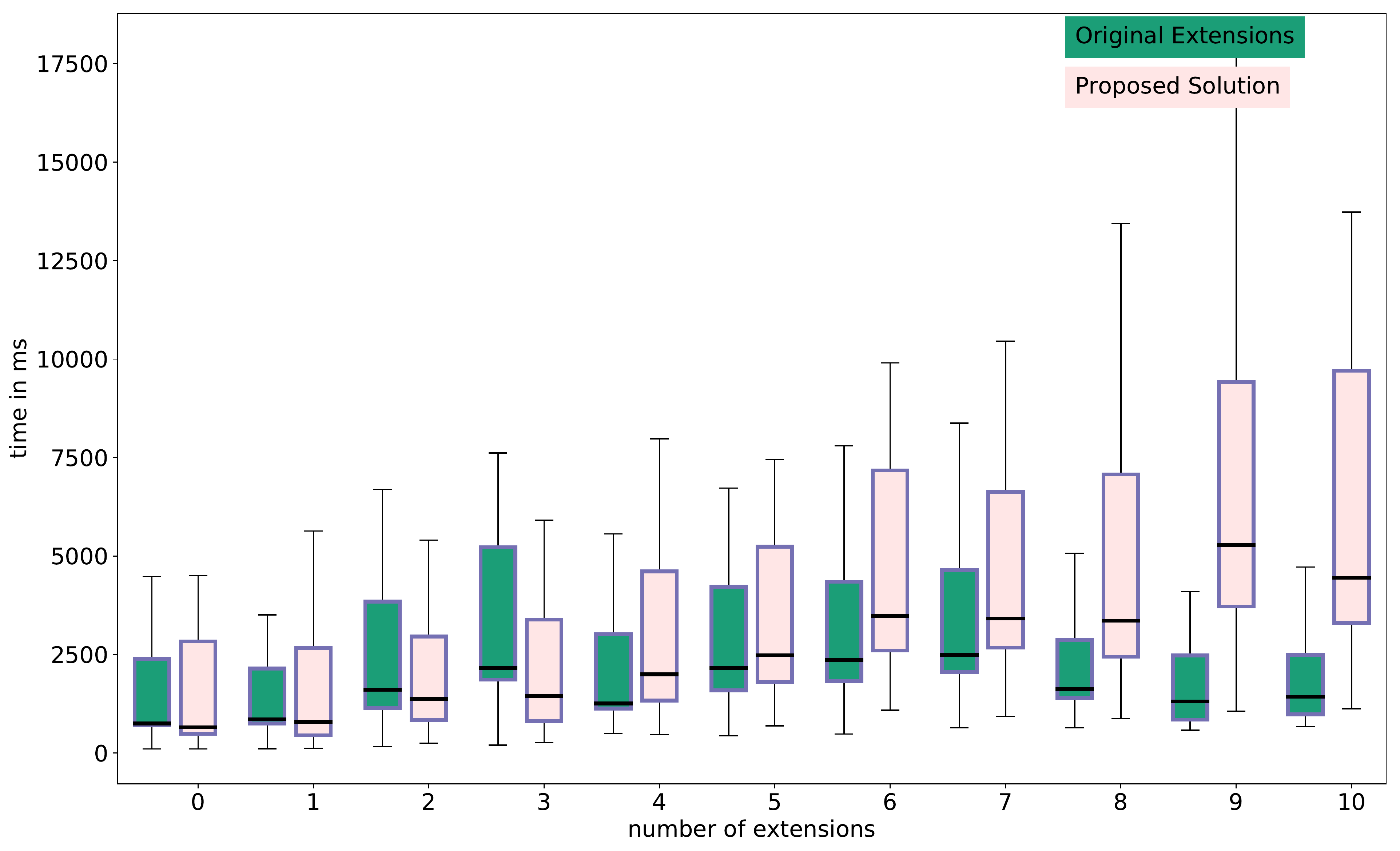}
		}
	}
	\subfloat[{\small Total Time}\label{fig:total_time}]{%
		\resizebox{7.5cm}{!}{
			\includegraphics[]{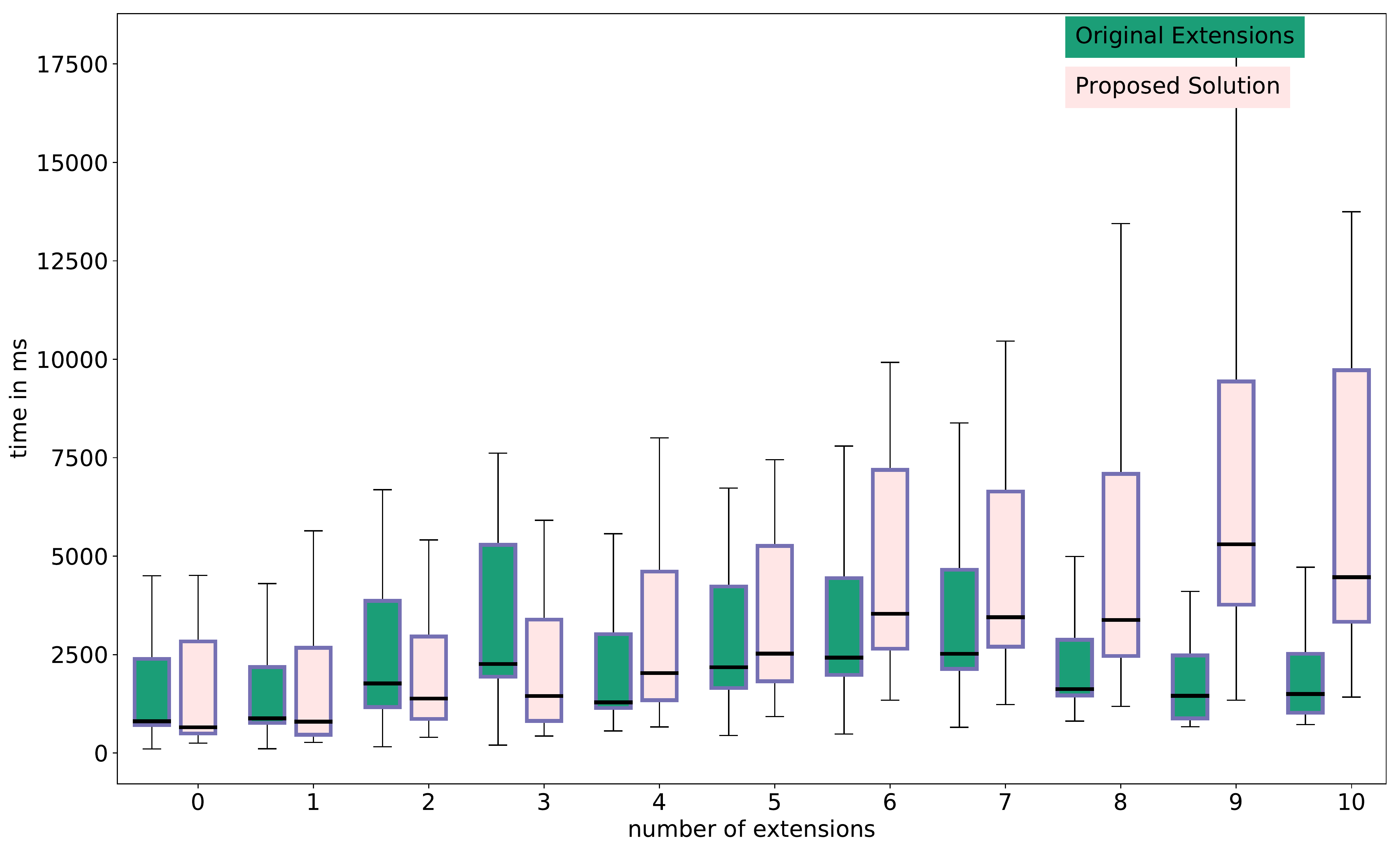}
		}
	}
	\caption{Evaluation of our proposal according to W3C parameters}
	\label{fig:w3ctimes}
\end{figure*}

Figure \ref{fig:htmlparse_time} shows the time that Chrome needs to parse the HTML. At this point, Chrome has already parsed the entire HTML file and creates the \gls*{DOM}. We can observe that, in the worst case (for 10 original extensions), our solution introduces a delay of 5000ms. Similarly, Figure \ref{fig:onload_time} shows the time needed for the browser to fire the event \textit{onLoad}. This event is critical because most of the extensions, jquery, and all libraries based on jquery wait for that event to be executed. From the results it is remarkable that the inclusion of our solution does not introduce undesired delays in the execution of this event in comparison to the default behavior.

The processing time measures when all resources have been loaded. Currently, the way the user knows when a given page has been totally loaded is when the spinner at the core of most browsers stops spinning. There are a bunch of external parameters that directly affect this time, such as the network overhead or the number of resources previously stored in the cache, among many others. All in all, we can conclude that the number of installed extensions has a potentially large impact on performance and, therefore, in the processing time as it is depicted in Figure \ref{fig:processing_time}. This, however, is only relatively critical as the average number of installed extensions is very low for most users.

Finally, Figure \ref{fig:total_time} shows the time needed to load the whole web page. The total time is calculated from the sum of processing and load times. This plot, together with the ones discussed before, show that content scripts of browser extensions are not totally decoupled from the rendering process and, therefore, they directly impact performance and user experience. 


In general, our solution increases very moderately the amount of time Chrome needs to render the content. This problem might be solved by modifying the browser's source code.

\section{Conclusions}\label{sec:conclusions}

In this paper we have discussed one important security and privacy implication of Chromium's extension model: the effects of one extension are visible to others in the execution pipeline. This can be exploited by a malicious extension that can, for example, get access to sensitive information or manipulate the \gls*{DOM} elements introduced by other extensions. We call this a {\em usual} attacker, in contrast to a {\em strong} attacker who has access to the effect of each single extension in the execution pipeline. A strong attacker may, in particular, install itself as the last extension in the pipeline and produce many copies interleaving itself in between all other extensions. In this way, it could be possible to get to know what all other extensions are doing and exploit this fact. We have shown examples on how to perform both a usual and a strong attack.

We have provided a proof-of-concept to address this problem which relies on replacing the pipeline execution model by one in which each extension executes in isolation, and then combine all individual effects to create the final \gls*{DOM}. Our implementation does this through a set of {\em monitor extensions}. As a first approach we decide to take the effect of the last extension in the pipeline. We could, however, easily provide a solution based on user intervention (asking the user to decide) or to apply a different policy (choose the first one, or non deterministically).  A more refined way to do so is left as future work (\eg\ one could gather information on how harmful the effects are, rank them and choose the less harmful using machine learning algorithms). 

\section*{Acknowledgments}
This work was partially supported by the the Swedish Research Council ({\it Vetenskapsr\aa det}) through the grant PolUser (2015-04154), the Swedish funding agency SSF through the grant Data Driven Secure Business Intelligence, the Spanish Goverment through MINECO grant SMOG-DEV  (TIN2016-79095-C2-2-R) and by the Regional Goverment of Madrid through the grant CIBERDINE (S2013/ICE-3095).

\bibliographystyle{plain}

\bibliography{main}

\end{document}